\documentclass[letterpaper]{article}
\usepackage{opex3}

\usepackage{upgreek, graphicx, amsmath, amssymb}

\newcommand{\fref}[1]{Fig.~\ref{#1}}
\newcommand{\Fref}[1]{Fig.~\ref{#1}}

\newcommand{\Sref}[1]{Section~\ref{#1}}

\newcommand{\ie}{\emph{i.e.}}

\newcommand{\rb}{$^{85}$Rb}

\newcommand{\name}{$\Lambda$MOT}

\makeatletter
\newlength \figwidth
\makeatother

\begin{document}

\title{Magneto-optical trapping and background-free imaging for atoms near nanostructured surfaces}
\author{Hamid\,Ohadi, Matthew\,Himsworth, Andr\'e\,Xuereb, and Tim\,Freegarde}
\address{School of Physics and Astronomy, University of Southampton,\\Southampton SO17~1BJ, United Kingdom}
\email{hamid.ohadi@soton.ac.uk}

\date{\today}

\begin{abstract}
We demonstrate a combined magneto-optical trap and imaging system that is suitable for the investigation of cold atoms near surfaces. In particular, we are able to trap atoms close to optically scattering surfaces and to image them with an excellent signal-to-noise ratio. We also demonstrate a simple magneto-optical atom cloud launching method. We anticipate that this system will be useful for a range of experimental studies of novel atom-surface interactions and atom trap miniaturization.
\end{abstract}


\bibliographystyle{osajnl}

\section{Introduction and motivation}
Over the past two decades, several configurations for magneto-optical traps have been demonstrated~\cite{Raab1987,Shimizu1991,Emile1992,Lee1996,Reichel1999}. The starting point for most geometries has been the original, `6-beam', configuration~\cite{Raab1987}, where the atom trap is created in the intersection of three counterpropagating laser beams. Despite it having the advantage that the atoms can be trapped far from any surface, thereby reducing spurious scatter in the imaging of such a trap, one cannot easily use this configuration for investigations into atom--surface interactions, for precisely the same reason. Another, more recent, configuration is the so-called `mirror MOT'~\cite{Reichel1999}, where the trap is formed a short distance away from a mirror, which also serves to reduce the number of necessary incident laser beam paths to two. The major drawback of such a configuration is its reduced optical access, due to the oblique angle of the field coils with respect to the mirror. The presence of a reflecting surface close to the trap also presents a problem of an entirely different nature. If the object of one's investigation is to observe the interaction between atoms and surfaces structured at the $\upmu$m scale, for example hemispherical mirrors of the type investigated in~\cite{Coyle2001}, the signal from the atoms will almost certainly be lost due to unwanted scattering of light into the optical system. MOTs on the meso- and microscopic scale, in particular, have received some recent interest~\cite{Folman2000}, but the small atom numbers in such traps have so far hindered their imaging and characterisation~\cite{Pollock2009}. In this article we propose a modified configuration that we call the `\name' and implement an imaging system based on a two-stage excitation process~\cite{Nez1993}, which help us overcome each of these limitations and aid our exploration of different atom--surface interactions.
\par
This paper is structured as follows. The next section is devoted to the description and characterization of our trap geometry. We then discuss the mechanism behind our multilevel imaging system and show how it does indeed allow for practically background-free imaging of the atom trap. The subsequent section discusses surface loading by magneto-optic launching, which allows us to load atoms onto a surface with a three-dimensional range of motion. Finally, we conclude and summarize the main features of our system.

\section{The \name}\label{sec:MOT}
\subsection{Description}
\begin{figure}[t]
 \centering
    \includegraphics[width=\figwidth]{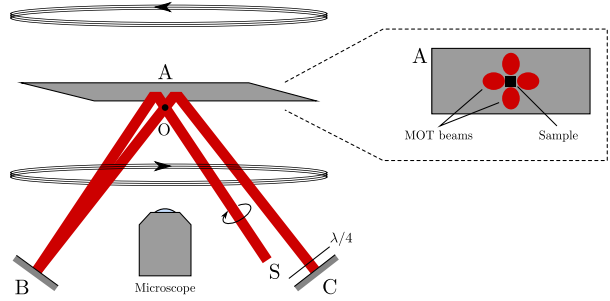}
\caption{(Color online.) Schematic of one of the two beam paths involved in our MOT geometry. S is the incoming beam; A, B, and C are mirrors. The component marked `$\lambda/4$' is a quarter-wave plate. The cold atom cloud forms in the intersection region, O. In this diagram we do not show a second, identical, beam, which provides trapping and cooling forces in the plane normal to the paper. The area of mirror A immediately adjacent to the trapped atoms is not illuminated, and can therefore be patterned or structured to explore atom--surface interactions. \emph{Inset:} The lower surface of mirror A, showing the MOT beams and the sample area, which is not illuminated by any of the beams.}
 \label{fig:VWMOT}
\end{figure}
\begin{figure}[t]
 \centering
    \includegraphics[width=\figwidth]{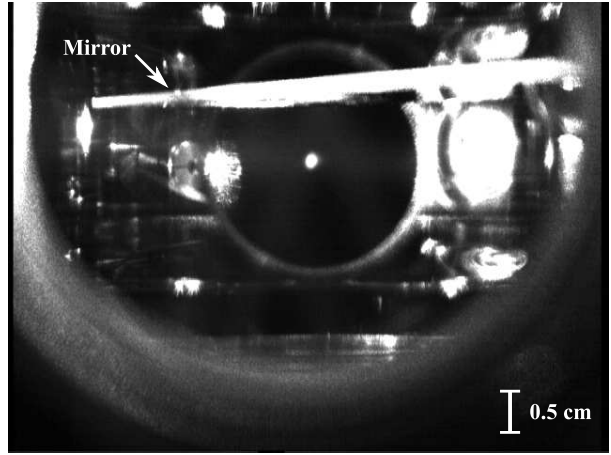}
\caption{Image of our MOT in operation, corresponding to \fref{fig:VWMOT}; mirror A is indicated in the picture.}
 \label{fig:MOTPic}
\end{figure}
A single beam of circularly polarized light of the right helicity is split using a non-polarizing beamsplitter, to generate the two beams that produce the trap, and a half-wave plate is inserted in one of the two resulting beams to achieve the correct polarizations. Each of these beams, denoted S, is then used to construct the geometry shown in \fref{fig:VWMOT}. Mirror C is set up so as to retroreflect the beam. Mirrors B and C, together with the quarter-wave plate, allow us to change the polarization in the retroreflected branch independently of the incoming polarization. In a normal mirror MOT, the polarizations cannot be modified independently of each other and the quadrupole axis has to be at an oblique angle to the mirror. The four beams traveling towards O thus have the correct polarizations to produce the trapping and cooling forces necessary to form a MOT in this plane. Combined with the second set of beams, this means that the MOT is formed in the intersection region of four pairs of counterpropagating beams. We note that alignment of mirror B such that the beam is retroreflected perfectly will recover the traditional mirror MOT beam geometry, albeit with the incorrect polarizations for a MOT cloud to form.
\par
Several advantages are apparent in the use of this geometry. The trapping volume is the entire overlap of the trapping beams, unlike that in a mirror MOT where half the trapping volume is rendered inaccessible by the presence of the mirror. Optical access is also much improved, both because the coils are oriented in such a way as to be less obstructive, and because we have removed the necessity of having a beam traveling in a plane parallel to mirror A. This allows us to use as much of the $360$\textdegree{} viewing angle in that plane as is necessary for imaging or manipulation beams. If this is not a requirement, a simpler set-up can alternatively be used, where only one set of beams is used in the double-`$\Lambda$' geometry, the trapping and cooling forces in the plane normal to the paper in \fref{fig:VWMOT} being produced by means of a separate pair of counterpropagating beams.
\\
An important advantage of this geometry is that the the double-`$\Lambda$' shape of the MOT beams affords better imaging of the trap, allowing microscope objectives to be mounted very close to it. With a custom-made objective, we can achieve high-NA imaging ($\mathrm{NA}>0.5$) and a diffraction-limited resolution of $<2$\,$\upmu$m. While a similar degree of optical access may be possible in the traditional 6-beam configuration, we note that this latter configuration is unsuitable for atom--surface interaction studies. In contrast, mirror A in our geometry can be replaced by any other suitable reflecting surface. One candidate for such a reflecting surface would be one of the surfaces of a Dove prism, which could then be used to form a two-dimensional bichromatic evanescent-field trap~\cite{Ovchinnikov1991} close to the mirror surface. This trap would be loaded from the MOT cloud using such techniques as magneto-optic launching, which is explained in \Sref{sec:Launching}.
\par
Aside from this marked increase in optical access, our system is simple to set up and operate. In particular, it requires fewer beam paths than a traditional MOT (two rather than three) and alignment of the beams is also easy: a CCD camera looking up at the mirror can be used to align the beams coarsely; once this is done, optimization of the cold atom signal provides the fine-tuning of the alignment.

\subsection{Characterization}
A typical trap, as shown in \fref{fig:MOTPic}, contains around $4\times 10^4$ \rb{} atoms and has a $1/e$ diameter of the order of $400$\,$\upmu$m along the minor axes. Combined with a measured trap lifetime $\tau_0\approx 6$\,s, this allows us to infer the trap loading rate, $N_0/\tau_0\approx 6.7\times10^3$\,s$^{-1}$. We measured a cloud temperature of $110\pm 40$\,$\upmu$K, the large uncertainty being due to the imprecision in measuring the cloud size.
\par
Typical parameters for the operation of our trap are: a detuning of $-14.9$\,MHz, or $-2.5$\,$\Gamma$ ($\Gamma\approx 6.1$\,MHz~\cite{Schultz2008}), for the cooling laser and a power of $6$\,mW divided between the two trapping beams (beam diameter: $6$\,mm). The minimum power necessary to produce the MOT was found to be $\approx\!1.3$\,mW in each of the two beams. The trap was loaded from background gas of a natural isotopic mixture of rubidium at a pressure of $10^{-9}$\,mbar. The cooling and repump lasers were locked using the DAVLL technique~\cite{Corwin1998} for long-term stability and flexibility of operation.

\section{Multilevel imaging system}\label{sec:Imaging}
\begin{figure}[t]
 \centering
    \includegraphics[width=0.7\figwidth]{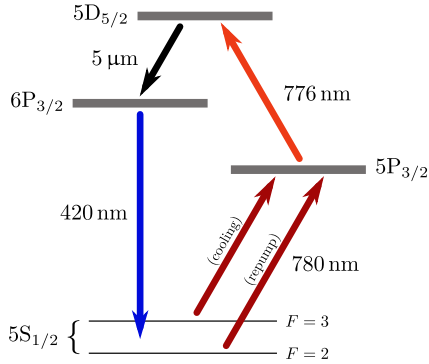}
\caption{(Color online.) The four-level system in \rb{} that we use to image our atoms. The MOT lasers ($780$\,nm) and a laser at $776$\,nm are used to induce a ladder transition. The population decays back to the ground state, via an intermediate state, and emits a $420$\,nm photon in the process. The hyperfine splitting of the excited states is not drawn for clarity.}
 \label{fig:420_Levels}
\end{figure}
The most common method of imaging a cold atom cloud in a MOT is fluorescence imaging. When the cloud is close to a reflecting surface both the cloud and its reflections will be seen by the imaging system (see Ref.~\cite{Clifford2001}, for example). This situation is exacerbated by the presence of surfaces that reflect unwanted light into the imaging optics and thereby decreasing the signal-to-noise ratio of the imaging system. \Fref{fig:MOTPic}, shows an example of the mirror in our system scattering the MOT beams into the imaging system.
\\
This problem may be overcome using two-stage excitation imaging. We make use of a four-level system in \rb{} (see \fref{fig:420_Levels} for details), similarly to Refs.~\cite{Sheludko2008} and~\cite{Vernier2009}; atoms in the $5$S$_{1/2}$ ground state are pumped to the $5$D$_{5/2}$ state via $780$\,nm and $776$\,nm radiation, the former being provided by one of the MOT beams, and then decay back to the ground state via an intermediate $6$P$_{3/2}$ state, emitting $420$\,nm radiation, which we detect. We note that a very similar system was recently used to produce a multiphoton MOT~\cite{SWu2009}. In our system, this process gives a significantly smaller signal than can be obtained through $780$\,nm fluorescence imaging. However, it has the benefit of being entirely background-free: in a well-shielded system, the entire $420$\,nm signal reaching the detector has its origin in the cold atom cloud. Off-the-shelf filters can then be used to remove the $780$\,nm radiation reaching the detector.

\subsection*{Generation of the $776$\,nm beam}
\begin{figure}[t]
 \centering
    \includegraphics[width=\figwidth]{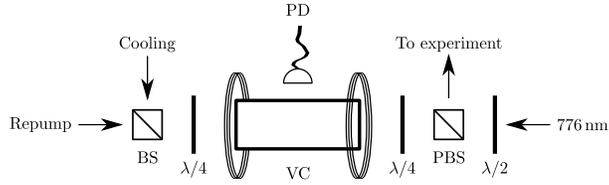}
\caption{$776$\,nm spectroscopy and locking system. (P)BS: (polarizing) beam splitter cube; $\lambda/4$: quarter-wave plate; $\lambda/2$: half-wave plate; VC: heated vapor cell; PD: filtered photodiode.}
 \label{fig:776Locking}
\end{figure}
The $776$\,nm beam is produced using a Sanyo DL7140-201S diode and the same external cavity diode laser design used to produce the MOT cooling and trapping beams. Since \rb{} has no spectral features in this wavelength range that are suitable for locking the laser frequency, a multilevel locking system is used (see \fref{fig:776Locking}). $5$\,mW from each of the MOT cooling and repump beams ($\approx\!780$\,nm) and $1.5$\,mW from the $776$\,nm beam, all rendered circularly polarized by the quarter-wave plates, enter the heated vapor cell (VC) from opposite ends. A large-area UV-enhanced filtered silicon photodiode (PD), operating in photovoltaic mode, picks up the resulting Doppler-free fluorescence and is amplified by means of a LMP7721 amplifier chip. Magnetic coils surrounding the heated vapor cell control the Zeeman shift of the magnetic sublevels of the atoms inside the cell, shifting this signal, and therefore the lock point, as required. Around $4$\,mW of the $776$\,nm beam is then mixed in with the MOT cooling and repump beams and sent through a fiber to the MOT.
\\
\begin{figure}[t]
 \centering
    \includegraphics[width=\figwidth]{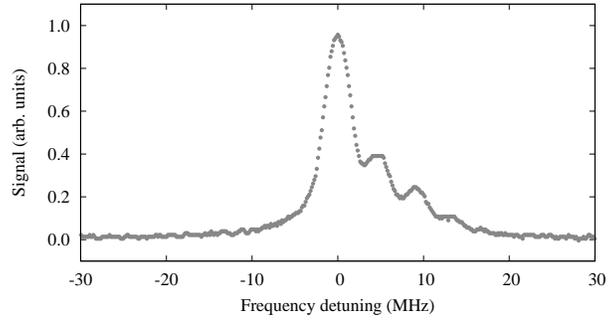}
\caption{$420$\,nm fluorescence from the vapor cell, observed on PD (see \fref{fig:776Locking}) as a function of the detuning of the $776$\,nm beam, with the cooling and repump beams locked and shifted by $80$\,MHz with respect to the frequencies required to make a MOT. The various peaks are due the hyperfine structure in \rb{}. To obtain these data, we removed the quarter-wave plates on either end of the vapor cell, thus having linearly polarized light entering the cell from both ends.}
 \label{fig:420Spectrum}
\end{figure}
\begin{figure}[t]
 \centering
    \includegraphics[width=\figwidth]{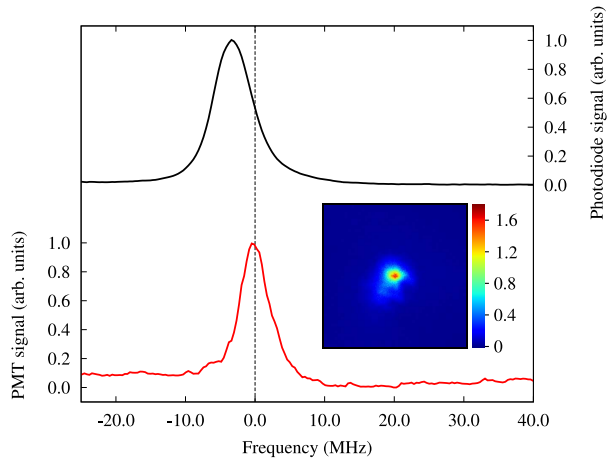}
\caption{(Color online.) $420$\,nm fluorescence observed on PD (solid black line, see \fref{fig:776Locking}) and on a PMT imaging the MOT cloud (solid red line) as a function of the detuning of the $776$\,nm beam. The zero on the frequency axis corresponds to the point at which the signal from the MOT cloud is highest; we lock to this point. The magnitude and sign of the shift between the two curves can be set arbitrarily by varying the magnetic field generated by the coils around the vapor cell. \emph{Inset:} MOT cloud imaged at $420$\,nm (scale in $10^3$ counts per second). This image is naturally background-free.}
 \label{fig:420nmPDPMT}
\end{figure}
\begin{figure}[t]
 \centering
    \includegraphics[width=\linewidth]{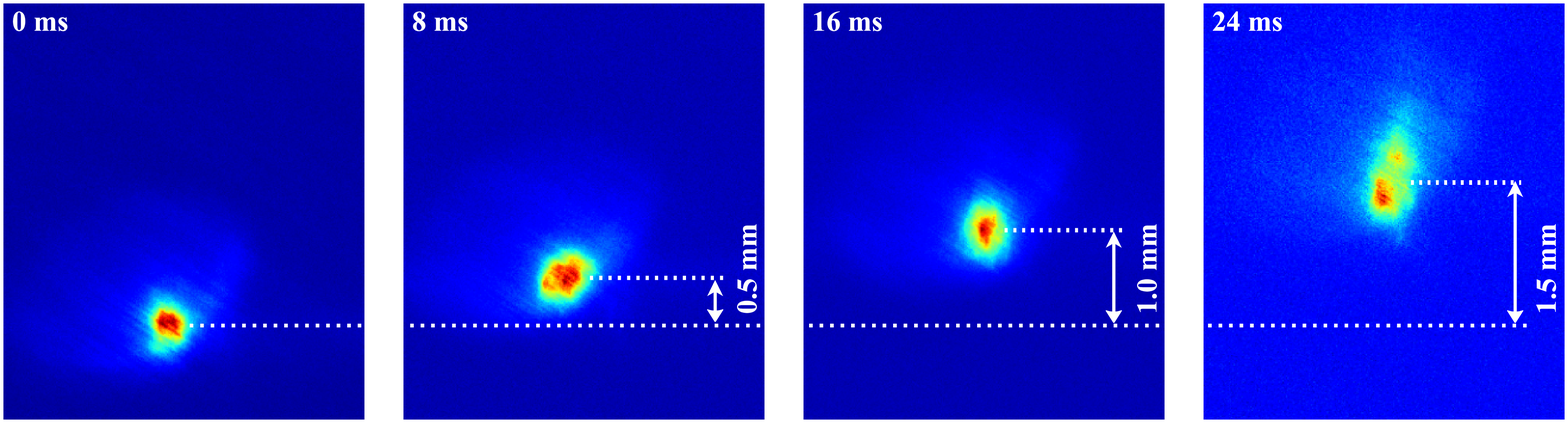}
\caption{(Color online.) A sequence of four false color fluorescence images, taken at $8$\,ms intervals, of the cloud before and after it has been given a magnetic impulse. The first shot (leftmost picture) shows the cloud just before the magnetic field is pulsed. The second, and subsequent, shots show the cloud at later times. The transfer efficiency after $24$\,ms is over $40$\%.}
 \label{fig:MagLaunch}
\end{figure}
We show a sample $420$\,nm signal, as detected at the photodiode, in \fref{fig:420Spectrum}, where the hyperfine splitting of the $5$D$_{5/2}$ level in \rb{} is evident in the shoulders on the right-hand side of the peak in the figure. The $776$\,nm laser diode is locked to the side of one the main peak, at the point indicated by the dashed line in \fref{fig:420nmPDPMT}, using a conventional PID circuit. The lock point is found by manually and slowly adjusting the frequency offset of the $776$\,nm laser to maximize the fluorescence from the MOT cloud. Evident in this latter figure are two well-resolved peaks, caused by the Autler--Townes splitting~\cite{Autler1955}. Locking at a detuning of around $6.5$\,MHz from the peak of absorption in the vapor cell gives the strongest signal in the MOT cloud, as recorded by the photomultiplier tube trace shown in the same figure.

\section{Surface loading by magneto-optic launching}\label{sec:Launching}
Transporting cold atoms from the region where the trap naturally forms to the sample is an essential part of many experiments investigating atom--surface effects. Several methods have been devised for moving cold atom clouds, including the use of push beams~\cite{Wohlleben2001} and moving magnetic coils~\cite{Lewandowski2003}. Push beams are easy to set up, requiring either the addition of one extra beam or the switching off of one of the counterpropagating beams, but cannot be used to push atom clouds towards highly reflective surfaces. Using moving magnetic coils requires a rather involved mechanical setup.
\par
We make use of a third method, which we call magneto-optic launching, for transport of the atom cloud by rapidly moving the trap center and then releasing the cloud, thereby imparting momentum to it. An auxiliary coil is added to the system in~\fref{fig:VWMOT}, above the upper MOT coil. After the MOT cloud forms, a long current pulse is applied to this auxiliary coil, which launches the cloud upward with a speed determined by the size and duration of the current pulse, and then the cloud is released from the trap by switching off the MOT beams after $20$\,ms. \Fref{fig:MagLaunch} shows a series of photographs of the cloud after being launched by a magnetic pulse. It can be seen that the pulse results in an approximately uniform vertical cloud speed of $0.063$\,m\,s$^{-1}$. The physical orientation of our system, with the mirror and sample being \emph{above} the trapping region, allow us to launch the cloud upwards with a much greater degree of control than would be possible if the cloud were merely dropped downwards.
\par
Finally, we note that the equilibrium distance of the MOT cloud from the mirror surface depends on the beam diameter and the size of the `sample area', \ie, the section of the mirror that acts as a sample and is not usable as a plane mirror. With a sample area diameter of $2$\,mm and beam diameter of $4$\,mm, the cloud can be made to form less than $4$\,mm away from the surface, allowing us to use the magneto-optic launching method to move the atoms closer to the surface for interaction studies.

\section{Conclusion}
We have introduced and characterized a modified magneto-optical trap geometry that allows the behavior of atoms close to surfaces to be explored with greater flexibility and better optical access than the standard configurations. A multilevel imaging system, which proves to be important in eliminating background signals and unwanted scatter when atoms are close to highly reflecting templated surfaces, was also characterized and explored. The combined system is therefore ideal for exploring the miniaturization of atom traps and is easily applied to a wide range of experiments.

\section*{Acknowledgments}
This work was supported by the UK EPSRC grants EP/E039839/1 and EP/E058949/1 and by the \emph{Cavity-Mediated Molecular Cooling} collaboration within the the EuroQUAM programme of the ESF.

\end{document}